\documentstyle[11pt,epsfig,moriond,amssymb]{article}

\def\Journal#1#2#3#4{{#1} {\bf #2}, #3 (#4)}

\def\mnras{{\em M.N.R.A.S.}}
\def\apj{{\em Ap. J.}}
\def\astlett{{\em Astr. Lett.}}

\newcommand\ba{\begin{array}}
\newcommand\ea{\end{array}}
\newcommand\bc{\begin{center}}
\newcommand\ec{\end{center}}
\newcommand\be{\begin{enumerate}}  
\newcommand\ee{\end{enumerate}}  
\newcommand\bi{\begin{itemize}}  
\newcommand\ei{\end{itemize}}  
\newcommand\bd{\begin{description}}  
\newcommand\ed{\end{description}}  
\newcommand\beq{\begin{equation}}  
\newcommand\eeq{\end{equation}}  
\newcommand\beqa{\begin{eqnarray}}  
\newcommand\eeqa{\end{eqnarray}}

\newcommand\eg{{\em e.g.,\ }}
\newcommand\etc{{\em etc}}

\newcommand\mathC{\mkern1mu\raise2.2pt\hbox{$\scriptscriptstyle|$}
                {\mkern-7mu\rm C}}

\newcommand\abs[1]{\vert {#1}\vert}

\renewcommand\l{\lambda}                        

\newcommand\th{\theta}

\newcommand\ph{\phi}

\newcommand\pdby[1]{{\frac{\partial}{\partial{ #1}}}}

\newcommand\pddby[1]{{\frac{\partial^2}{\partial{ #1}^2}}}

\newcommand\Y[2]{Y_{#1}(\th_{#2},\ph_{#2})}

\begin{document}
\vspace*{2cm}
\title{TOPOLOGICAL ANALYSIS OF HIGH--RESOLUTION CMB MAPS
\footnote{Full size, full colour images can be downloaded from {\it 
http://www.tac.dk/$\sim$wandelt}}}

\author{ B.D. WANDELT, E. HIVON, K.M. GORSKI, }

\address{Theoretical Astrophysics Centre (TAC), Juliane Maries Vej 30,\\
2100 Copenhagen $\O$, Denmark}

\maketitle\abstracts{ We report the development of numerical tools for the
topological analysis of sub--degree resolution, all--sky
maps. Software to be released in the HEALFAST (V0.9) package defines
neighbour relationships for the HEALPIX 
tessellation of the sphere. We apply this routine to a fast extrema
search which scales strictly linearly in the number of pixels, $N_{p}$. 
We also present a highly efficient algorithm for 
simulating the gradient vector and
curvature tensor fields ``on--the--fly'' with the temperature map,
needing only of order
$N_{p} \log_2 N_{p}$ more operations.}

\section{Introduction}\label{sec:intro}

The Microwave Anisotropy Probe (MAP) and the Planck Surveyor satellite
missions promise to produce high--resolution full sky maps of the
Cosmic Microwave Background (CMB) within
the next decade. These maps will contain a great deal of
cosmologically relevant information and therefore present a
tremendous opportunity for doing
high--precision cosmology. Owing to the unprecedented wealth of data,
this opportunity brings with it the challenge 
of developing tools which help
extract this information in manageable processing time. 

Much attention has been given to the problem
of estimating the angular power--spectrum of CMB
anisotropies. To illustrate the difficulties one is facing, we note that even
choosing the fastest presently
available algorithm and allowing for parallelisation and technological
advance, the estimation of the angular power--spectrum $C_l$ of CMB
anisotropies will take $\gtrsim 6$ years \cite{borrill} for future 
million pixel maps.

While the $C_l$ spectrum is the statistic of choice for a large class
of cosmological models where the primordial fluctuations are Gaussian
distributed, the possibility remains that another physical 
mechanism produced structure in the universe. In this case the CMB
maps will almost certainly be non--Gaussian and contain more
information than merely the power on different angular scales. Failing this,  
non--linear gravitational effects and foregrounds will produce detectable
non--Gaussian signals. 

Even for Gaussian maps, it has been shown that extrema statistics
capture information about cosmological parameters and their theory is
well developed \cite{sazhin,BE,VJ,BSMCS}. The importance of studying
these and other topological properties 
of CMB temperature and polarisation
has been noted by several authors \cite{topos}.
Extrema have
the advantage that they dominate the noise and are therefore very
robust. As a consequence they have been among the first scientific
results reported from recent CMB observations (one of them even at
this conference \cite{spot}!).

The aim of this talk is to present tools which allow the
implementation of a wide 
range of data analysis and processing techniques on the sphere.
\newpage
\section{Tools}
\subsection{Neighbour relationships on the sphere}
Finding neighbours of a given pixel is essential for performing
many local operations on digitised data, such as finding
extrema, saddle points and zero crossings, finite differencing, 
extraction of patches, real--space filtering {\etc}. 
While next--neighbour relationships are
trivially defined on a lattice in the plane, 
on the sphere it is necessary to match distinct coordinate
patches. Further, in a spherical topology the number of next--neighbours is
not the same for all pixels. 
\begin{figure}
\centerline{\psfig{figure=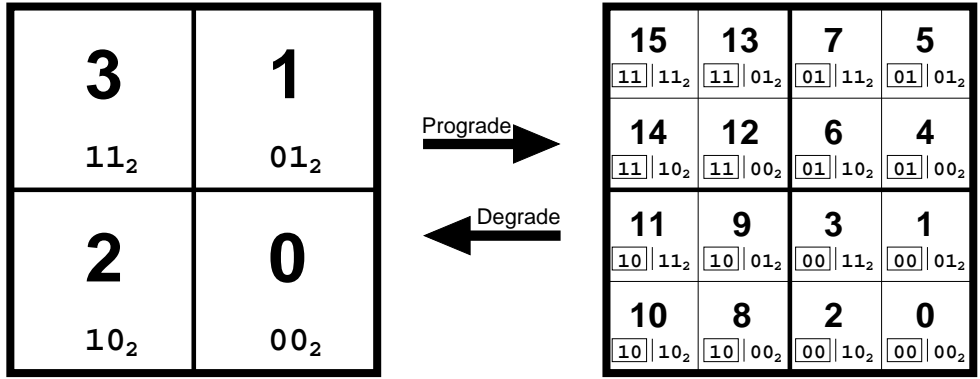,height=2in}}
\caption
{Quadtree pixel numbering. The coarsely pixelised coordinate patch on
the left consists 
of four pixels. Two bits suffice to label the pixels. 
To increase the resolution, every 
pixel splits into 
4 daughter pixels shown on the right. These daughters inherit the pixel
number of their 
parent (boxed) and acquire 
two new bits to give the new pixel number.
Several such coordinate patches are joined at the boundaries to cover
the sphere. All pixels carry a prefix (here omitted for clarity) 
which identifies which patch they belong to.
\label{fig:quadtree}}
\end{figure}

We base our methods on the HEALPIX tessellation of the sphere
\cite{gorski}. However, even when a pixelisation of the sphere is
given, we must still decide 
how to number the pixels. We choose the hierarchical, quadtree numbering
scheme illustrated in Figure \ref{fig:quadtree}. This scheme is
well--known and was used for the COBE pixelisation. 
Its main advantages
for finding next--neighbours are that 
\bi
\item a neighbour can be found by direct operation on the binary
representation of the pixel number which is computationally very cheap
(finding all next--neighbours of $12\times 10^6$ pixels takes less than 10
seconds \footnote{All timings are ``wall clock''
for a single SGI R10000 194MHz processor.});
\item the algorithm is manifestly resolution independent owing to the
self--similarity of the numbering scheme;
\item once an algorithm exists that finds next--neighbours,
it can easily be extended to find a whole neighbourhood of a pixel by
first degrading, then finding next--neighbours and then
prograding.
\ei

A disadvantage is that in its implementation the boundary matching of
coordinate patches has to be customised for the chosen pixelisation
scheme. But even here the symmetries of the quadtree are useful. For
example, if two patches like the one on the right of Figure
\ref{fig:quadtree} were joined side by side, neighbours of pixels at
the joining edge can be found by inverting all even bits of their
pixel number, \eg $4={\sf \underline{0}1\underline{0}0_2}
\leftrightarrow 14={\sf \underline{1}1\underline{1}0_2}$. This and
analogous prescriptions generalise to arbitrary resolution.

\begin{figure}
\vskip -2.1in
\psfig{figure=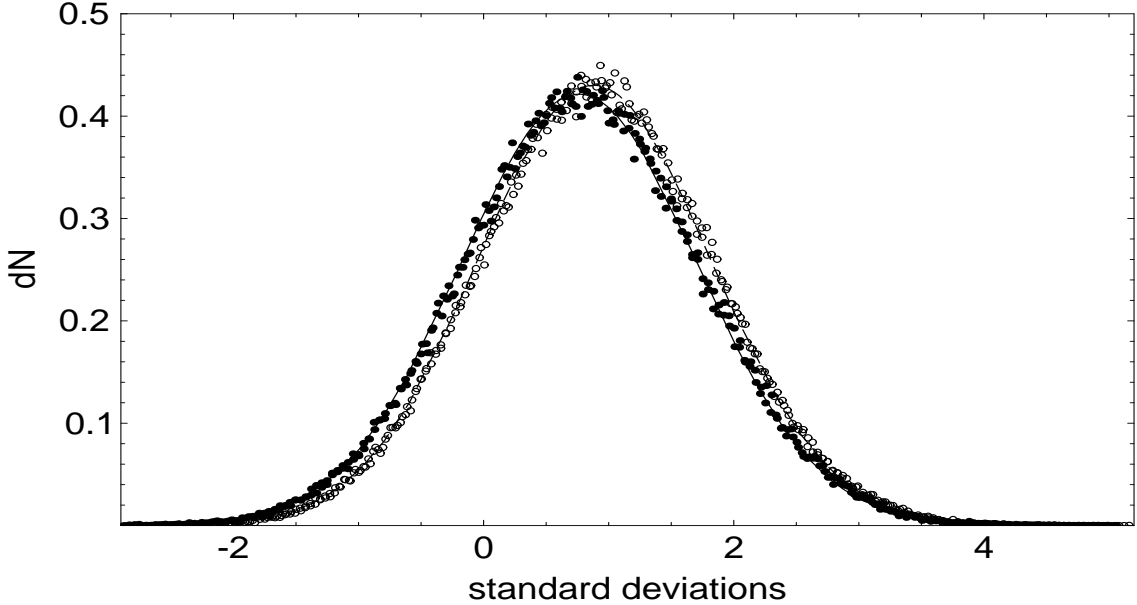,height=7in,width=6in}
\vskip -2.1in
\caption{Probability density for finding a maximum (minimum) above
(below) a certain excursion level. The minima distributions have been
flipped over to coincide with the maxima distributions. Open dots are 
found from one realization of an $\Omega=.2$ map, 
filled dots are for standard CDM. 
Lines show the  theoretically expected distributions (solid: SCDM, dashed: OCDM).
\label{fig:extrema}}
\end{figure}

As an example application, we show in Figure \ref{fig:extrema} 
the theoretical prediction and our
numerical results for an extrema search in a ``standard'' CDM 
and an open model 
with $\Omega=0.2$  ($H_0=50$km/s/Mpc,
$\Omega_b=0.05$ for both, $\th_{FWHM}=10'$ and zero noise). 
While it is clear that these probability
distributions contain much less information than the anisotropy
spectrum (they form a 3 parameter family if the total number of
pixels is counted), the distributions shown clearly
distinguish the two models. We stress that this is a realistic
situation, since we used data only from one map. 

This example do not even begin to explore the possibilities of
these methods. By just measuring the probability distribution of
excursion of an extremum from the mean, we are
throwing away valuable scale and phase information. A more careful
analysis will reveal more about the specific advantages
of topological analyses of CMB maps.

\subsection{Map simulation and computation of derivative fields}
\begin{figure}
\centerline{\psfig{figure=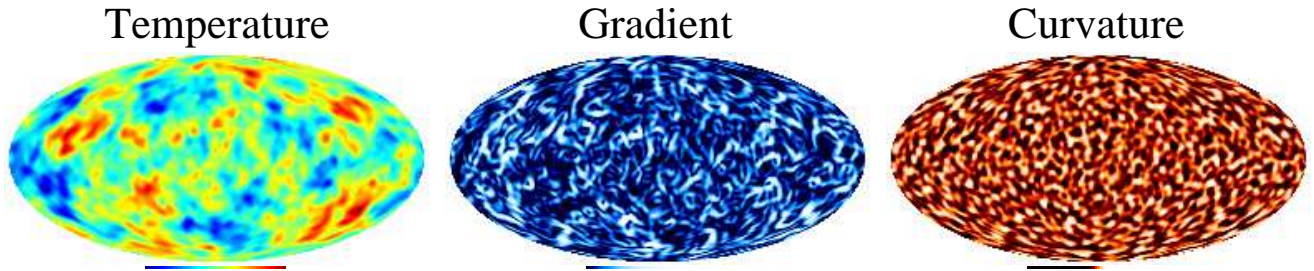,height=1.5in}}
\vskip -.2in
\caption{From left to right: the temperature, the square of the gradient
vector and the trace of the curvature tensor for a standard CDM sky
smoothed at 5 degrees for easier visualisation.
\label{fig:simulation}}
\end{figure}

The derivative fields of a map are  
important  for the study of
the properties of its stationary points. A Fast Spherical Harmonics
Transform code exists for 
the synthesis of Gaussian HEALPIX temperature maps with a given $C_l$
spectrum \cite{HG}. A priori, generating the derivative fields would
take just as long for each component as the map generation itself.
The time limiting factor for map generation is the
calculation of the associated Legendre polynomials $P^m_l(\cos\theta)$
for each ring of constant co--latitude $\theta$. We exploit this
fact by calculating the derivative fields at the same time as the
temperature map and exploiting the following differential relations:
\beq
\ba{rcl}
\pdby{\ph}\Y{lm}{} &=& im \Y{lm}{}\\
\pdby{\th}\Y{lm}{} &=& -m \cot\th\Y{lm}{} 
    -\sqrt{l(l+1)-m(m-1)}\Y{lm-1}{}e^{i\ph}\\
\pddby{\th}\Y{lm}{} &=& -\left(\left[l(l+1)-\frac{m^2}{\sin^2\th}\right]+
    \cot\th\pdby{\th}\right)\Y{lm}{}
\ea
\eeq
In this way, only five more Fast Fourier
Transforms have to be computed for each ring on the sphere, with the 
total added cost scaling as 
$N_p\log_2 N_{p}$. For a map with $N_p=12.6\times10^6$ pixels, this
amounts to an added 15 minutes of wall clock computation time for all
five derivative components compared with 20 minutes for the
temperature alone. In Figure
\ref{fig:simulation} we display temperature anisotropies, the square
norm of the gradient vector
$\abs{\nabla_i T}^2=\frac1{sin^2\th}(\pdby{\ph}T)^2+(\pdby{\th}T)^2$ 
and the trace of the
curvature tensor ${\rm Tr}(\l_{ij})=
\frac1{sin^2\th}\pddby{\ph}T+\pddby{\th}T$, for the standard CDM model.

\section{Conclusions}
In this talk we presented tools which are generally applicable for the
analysis of topological properties of maps on the sphere such as the 
detection of extrema, saddle points and zero crossings of a
function, even though they can be used more generally for any kind of
local analysis of spherical data sets.
The computational time required for all sky computations scales
strictly as the number of 
pixels in the map, $N_{p}$. Further,
we extend existing simulation tools for Gaussian fields to generate
not only temperature maps but also the associated gradient vector and
the curvature tensor fields, with the extra computational cost scaling as 
$N_{p} \log_2 N_{p}$.

\section*{Acknowledgments}
This work was supported by Dansk Grundforskningsfond through its
funding for TAC.


\section*{References}
\begin{thebibliography}{99}
\bibitem{borrill} J. Borrill, preprint, astro-ph/9712121.

\bibitem{sazhin} M. Sazhin, \Journal{\mnras}{216}{25{\sc p}}{1985}.

\bibitem{BE} J.R. Bond and G. Efstathiou, \Journal{\mnras}{226}{655}{1987}.

\bibitem{VJ}N. Vittorio and R. Juszkiewicz, \Journal{\apj}{314}{L29}{1987}.

\bibitem{BSMCS}R.B. Barreiro {\it et al.}, \Journal{\apj}{478}{1}{1997}.

\bibitem{topos} M. Sazhin and Toporenskii
\Journal{\astlett}{22}{791}{1996}; P. Arbuzov {\it et al.},
\Journal{Int. J. Mod. Phys. D}{6}{409}{1997}; P. Arbuzov {\it et al.},
\Journal{Int. J. Mod. Phys. D}{6}{515}{1997}; P.D. Naselsky and
D.I. Novikov, preprint, astro-ph/9801285.

\bibitem{spot} J. Baker, in this volume.

\bibitem{gorski} K.M. Gorski, in preparation.

\bibitem{HG} E. Hivon and K.M. Gorski, in preparation.

\end{thebibliography}
\end{document}